\documentclass[journal=jacsat,manuscript=article]{achemso}
\usepackage{xr}
\makeatletter
\newcommand*{\addFileDependency}[1]{
  \typeout{(#1)}
  \@addtofilelist{#1}
  \IfFileExists{#1}{}{\typeout{No file #1.}}
}
\makeatother

\newcommand*{\myexternaldocument}[1]{
    \externaldocument{#1}
    \addFileDependency{#1.tex}
    \addFileDependency{#1.aux}
}

\myexternaldocument{Supportinginformation}

\usepackage[version=3]{mhchem} 
\usepackage[T1]{fontenc} 
\usepackage{gensymb}
\usepackage{xcolor}

\usepackage{CJKutf8} 
\newcommand{\zcite}[1]{\scalebox{1.5}[1.5]{\raisebox{-0.84ex}{\cite{#1}}}} 

\author{Yi Luo}
\affiliation[School of Electronics]
{Beijing Key Laboratory of Quantum Devices, Key Laboratory for the Physics and Chemistry of Nanodevices, and School of Electronics, Peking University, Beijing 100871, China}
\alsoaffiliation[School of Physics]
{Institute of Condensed Matter and Material Physics, School of Physics, Peking University, Beijing 100871, China}

\author{Xiao-Fei Liu}
\affiliation[BAQIS]
{Beijing Academy of Quantum Information Sciences, Beijing 100193, China}

\author{Zhi-Hai Liu}
\affiliation[BAQIS]
{Beijing Academy of Quantum Information Sciences, Beijing 100193, China}

\author{Weijie Li}
\affiliation[School of Electronics]
{Beijing Key Laboratory of Quantum Devices, Key Laboratory for the Physics and Chemistry of Nanodevices, and School of Electronics, Peking University, Beijing 100871, China}

\author{Shili Yan}
\affiliation[BAQIS]
{Beijing Academy of Quantum Information Sciences, Beijing 100193, China}

\author{Han Gao}
\affiliation[School of Electronics]
{Beijing Key Laboratory of Quantum Devices, Key Laboratory for the Physics and Chemistry of Nanodevices, and School of Electronics, Peking University, Beijing 100871, China}

\author{Haitian Su}
\affiliation[School of Electronics]
{Beijing Key Laboratory of Quantum Devices, Key Laboratory for the Physics and Chemistry of Nanodevices, and School of Electronics, Peking University, Beijing 100871, China}
\alsoaffiliation[School of Physics]
{Institute of Condensed Matter and Material Physics, School of Physics, Peking University, Beijing 100871, China}

\author{Dong Pan}
\email{pandong@semi.ac.cn}
\affiliation[CASIOS]
{State Key Laboratory of Superlattices and Microstructures, Institute of Semiconductors,Chinese Academy of Sciences, P.O. Box 912, Beijing 100083, China}

\author{Jianhua Zhao}
\affiliation[CASIOS]
{State Key Laboratory of Superlattices and Microstructures, Institute of Semiconductors,Chinese Academy of Sciences, P.O. Box 912, Beijing 100083, China}

\author{Ji-Yin Wang}
\email{wang_jy@baqis.ac.cn}
\affiliation[BAQIS]
{Beijing Academy of Quantum Information Sciences, Beijing 100193, China}

\author{H. Q. Xu}
\email{hqxu@pku.edu.cn}
\affiliation[School of Electronics]
{Beijing Key Laboratory of Quantum Devices, Key Laboratory for the Physics and Chemistry of Nanodevices, and School of Electronics, Peking University, Beijing 100871, China}
\alsoaffiliation[BAQIS]
{Beijing Academy of Quantum Information Sciences, Beijing 100193, China}

\title{One-dimensional quantum dot array integrated with charge sensors in an InAs nanowire }

\begin{document}

\begin{CJK}{UTF8}{gbsn} 

\newpage







\begin{abstract}
We report an experimental study of a one-dimensional quintuple-quantum-dot array integrated with two quantum dot charge sensors in an InAs nanowire. The device is studied by measuring double quantum dots formed consecutively in the array and corresponding charge stability diagrams are revealed with both direct current measurements and charge sensor signals. The one-dimensional quintuple-quantum-dot array are then tuned up and its charge configurations are fully mapped out with the two charge sensors. The energy level of each dot in the array can be controlled individually by using a compensated gate architecture (i.e., "virtual gate"). After that, four dots in the array are selected to form two double quantum dots and ultra strong inter-double-dot interaction is obtained. A theoretical simulation based on a 4-dimensional Hamiltonian confirms the strong coupling strength between the two double quantum dots. The highly controllable one-dimensional quantum dot array achieved in this work is expected to be valuable for employing InAs nanowires to construct advanced quantum hardware in the future.    
\end{abstract}


\newpage

Semiconductor nanowires with strong spin-orbit interactions have played important roles in pursuing a variety of quantum processors, including spin qubits\cite{Loss1998,Nadj_Perge2010}, gate tunable superconducting qubits\cite{Larsen2015,Lange2015}, Andreev spin qubits\cite{Chtchelkatchev2003,Padurariu2010,Hays2021} and Majorana-based topological qubits\cite{lutchyn2010,oreg2010,Aasen2016,Karzig2017}. Benefiting from high material quality, possibility to be transferred freely and compatibility with superconducting components, pioneer researches in the field of solid-state quantum computations have been accomplished on semiconductor nanowires. For instance, gate-tunable superconducting qubits or/and spin qubits have been demonstrated in InAs\cite{Nadj_Perge2010,Larsen2015,Lange2015,Casparis2016,Sabonis2020}, InSb\cite{Berg2013}, and Ge/Si core/shell nanowires\cite{Froning2021,Zhuo2023,Zheng2023}, and Andreev spin qubits with strong qubit-qubit coupling have been realized in InAs nanowires\cite{Marta2023,Marta2024}. Particularly, ultrafast spin qubit manipulations exceeding a few hundred MHz have been achieved in Ge/Si core/shell nanowires\cite{Froning2021} and Ge hut nanowires\cite{KeWang2022,Liu2023}. Aside from well-established qubits, nanowires with strong spin-orbit interactions are especially interesting in searching for Majorana zero modes\cite{lutchyn2010,oreg2010} and significant progress has been made on InAs and InSb nanowires\cite{Lutchyn2018,Mourik2012,Deng2012,Das2012,Deng2016,Wang2022,vanLoo2023}. Recently, researchers have successfully realized Kitaev chains composed of quantum dots and superconducting electrodes in InSb nanowires\cite{Dvir2023} and have demonstrated practical protections of zero energy modes from local electrochemical potential variations\cite{Bordin2024}. Thus, semiconductor nanowires have become versatile platforms, fueling the progress of solid-state quantum bits profoundly. 

Among semiconductor nanowires, InAs nanowires manifest themselves as an unique material which allows the study of all above qubit schemes. This arises from their special properties such as strong spin-orbit interactions\cite{Fasth2007spin_orbit,Wang2016APL,Wang2018Nanolett, Iorio2019Nanolett}, large Land\'e g factors\cite{Csonka2008gFactor,Wang2016APL} and compatibility with epitaxially in-situ growth of superconductors\cite{Krogstrup2015,Kanne2021,Pan2022}. In InAs nanowires, making quantum dots (QDs) is particularly important since qubit definition and quantum state readout normally rely on QDs in aforementioned qubit researches. In addition, coupled QDs in a linear array have shown to be highly tunable systems, which can serve as quantum simulators\cite{Hensgens2017,Kiczynski2022}. Significant research efforts have been poured into gate-defined QDs in InAs nanowires to investigate their charge and spin characteristics\cite{Fasth2007spin_orbit,Wang2016APL,Wang2018Nanolett,Iorio2019Nanolett,Csonka2008gFactor,Fasth2005,Pfund2007,Nadj_Perge2010PRB,Wang2017}. In spite of these efforts, InAs nanowires merely support a limited number of coupled QDs and meanwhile their charge state reading is normally difficult\cite{Mu2021Nanoscale}. This is mainly due to the challenge of integrating charge sensors on such a one-dimensional system. Thus, expanding the number of coupled QDs together with charge detectors in InAs nanowires is highly desired in order to target more complex quantum tasks. 

In this work, we report an experimental study of a one-dimensional quintuple-quantum-dot (QQD) array integrated with two charge sensors in an InAs nanowire. Both charge sensors exhibit sufficient sensitivities to discern the charge states of any QD in the array. Then, the QQD array is built up and the energy level of each dot can be tuned individually by means of the virtual gate control. After that, four QDs in the array are chosen to form two double quantum dots (DQDs) and strong coupling between the two DQDs is revealed. A numerical simulation based on a 4-dimensional Hamiltonian validates the strong coupling strength. We expect that the highly controllable one-dimensional QD array achieved in this work lays a foundation for executing more advanced quantum missions in nanowire systems in the long run.

\begin{figure*}[!t] 
\centering
\includegraphics[width=1\linewidth]{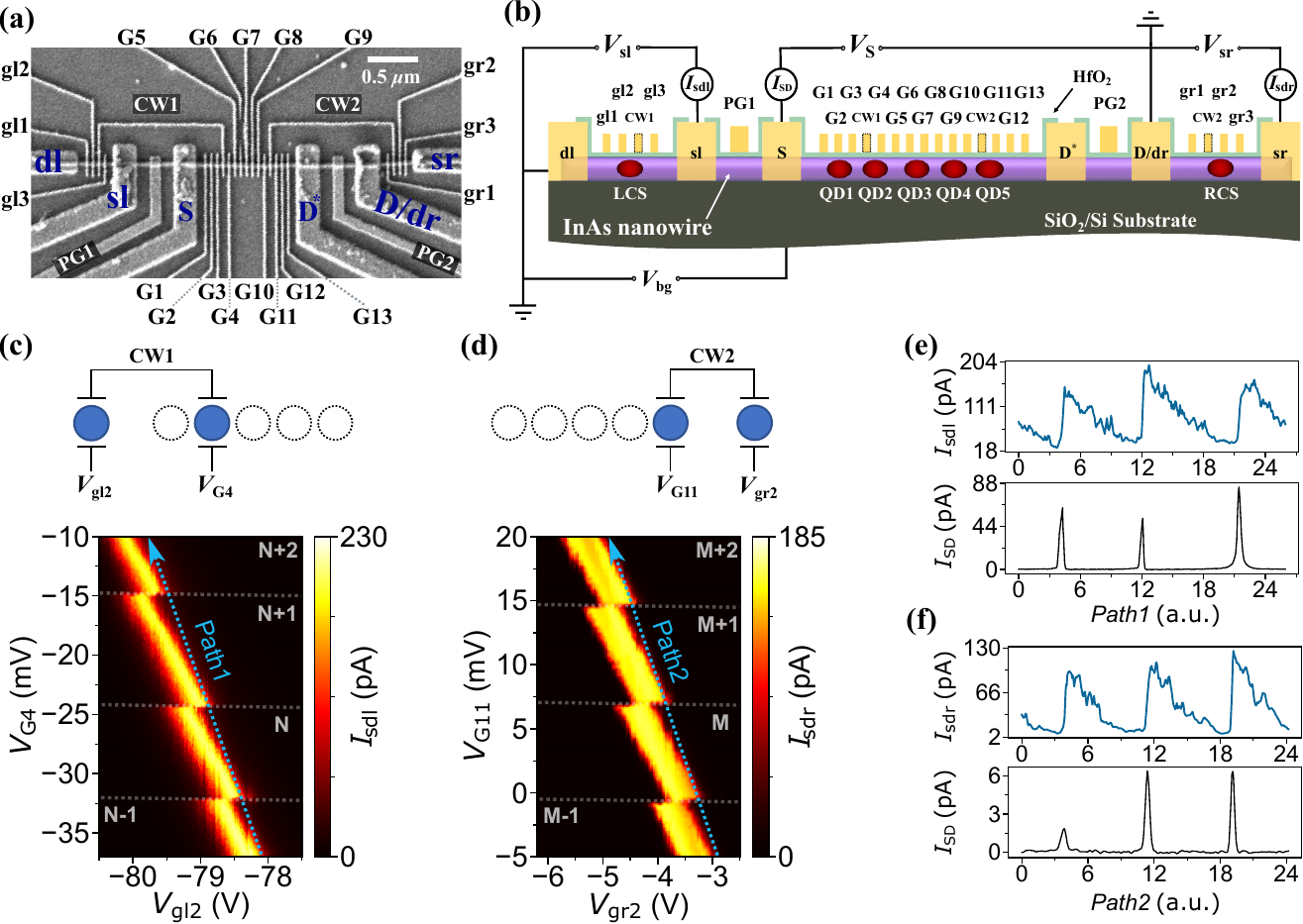}
\caption{\doublespacing \textbf{(a)} SEM image of the measured device. The diameter of the InAs nanowire is $\sim30\,\mathrm{nm}$. Contact electrodes (dl, sl, S, D*, D/dr and sr) are made from $5/70\,\mathrm{nm}$ Ti/Au. Fine gates (gl1-gl3, gr1-gr3 and G1-G13), metal coupling wires (CW1 and CW2) and pinch-off gates (PG1 and PG2) are made from $5/30\,\mathrm{nm}$ Ti/Au. \textbf{(b)} Cross-sectional schematic view of the device and measurement setup. The gates G1-G13 are used to control the QQD (QD1-QD5), the gates gl1-gl3 are for the left sensor dot (LCS) and the gates gr1-gr3 are for the right sensor dot (RCS). Current through LCS ($I_{sdl}$), QQD ($I_{SD}$) and RCS ($I_{sdr}$) are measured while applying bias voltage $V_{sl}$, $V_{S}$ and $V_{sr}$, respectively. Note that the gate G6 and the electrode D* are not functional. \textbf{(c)} $I_{sdl}$ as a function of $V_{G4}$ and $V_{gl2}$ when LCS and QD2 are tuned up (bottom panel), and corresponding dot configuration (top panel). \textbf{(d)} $I_{sdr}$ as a function of $V_{G11}$ and $V_{gr2}$ when RCS and QD5 are tuned up (bottom panel), and corresponding dot configuration (top panel). \textbf{(e)} The current $I_{sdl}$ and $I_{SD}$ as a function of the gates scanned along Path1 shown in figure (c). \textbf{(f)} The current $I_{sdr}$ and $I_{SD}$ as a function of the gates scanned along Path2 shown in figure (d).} \label{figure:1}
\end{figure*}

Figure \ref{figure:1}a shows a scanning electron microscope (SEM) image of the device comprising a QQD and two charge sensor dots. The device is fabricated from a single-crystalline pure-phase InAs nanowire grown by molecular-beam epitaxy (MBE)\cite{Pan2014NanoLett}. Contact electrodes (dl, sl, S, D*, D/dr and sr) are made from $5/70\,\mathrm{nm}$ Ti/Au. A 10-$\mathrm{nm}$-thick $\mathrm{HfO_\mathrm{2}}$ dielectric layer is grown subsequently by atomic layer deposition (ALD). Then, fine finger gates (gl1, gl2, gl3, gr1, gr2, gr3 and G1-G13), metal coupling wires (CW1 and CW2) and pinch-off gates (PG1 and PG2) are made on top of the nanowire with $5/30\,\mathrm{nm}$ Ti/Au. For fine finger gates and metal coupling wires, the line width is $\sim 20\,\mathrm{nm}$ and the pitch is $\sim 60\,\mathrm{nm}$. More details about device fabrications can be found in our previous relevant works\cite{Wang2018Nanolett,Mu2021Nanoscale,Li2023}. Figure \ref{figure:1}b displays a cross-sectional schematic of the device together with the measurement setup. As seen in the figure, the gates PG1 and PG2 are designed to cut the nanowire electrically into three segments, which are used to construct two sensors and a QD array afterwards. The gates gl1-gl3 are used to control the left charge sensor dot (LCS) in the left segment of the nanowire and the gates gr1-gr3 is for the right charge sensor dot (RCS) in the right segment of the nanowire. The gates gl1 and gl3 (gr1 and gr3) work as barrier gates to define LCS (RCS), while the gate gl2 (gr2) is a plunger gate to tune the electrochemical potential of LCS (RCS). The gates G1-G13 are used to control the QQD (QD1-QD5) in the middle segment of the nanowire. Two metal wires (CW1 and CW2) is to enhance the capacitive coupling between the two sensors and the QQD. Bias voltages $V_\mathrm{sl}$, $V_\mathrm{sr}$ and $V_\mathrm{S}$ are applied to LCS, RCS, and QQD, respectively, while corresponding current flow $I_\mathrm{sdl}$, $I_\mathrm{sdr}$ and $I_\mathrm{SD}$ are measured. As the lead D* is not functional, we use the electrode D/dr as a drain lead for both the QQD and RCS. The back gate with voltage $V_\mathrm{bg}$ is used to adjust the electrochemical potential of the nanowire globally. The device is measured in a $^3\mathrm{He/}^4\mathrm{He}$ dilution refrigerator at a base temperature of $\sim 20\,\mathrm{mK}$. In all measurements, $V_\mathrm{bg}$ is fixed at $2.5\,\mathrm{V}$ to open up the whole nanowire in the beginning, $V_\mathrm{PG1}$ is set at $-2\,\mathrm{V}$ to pinch off the part below the gate PG1, and $V_\mathrm{PG2}$ is at $0\,\mathrm{V}$ ensuring the current of the QQD to be drained via the lead D/dr.  

With the device, the pinch-off characteristics of all gates are initially explored and the threshold voltages are presented in Table S1 in the Supporting Information. After that, we start to investigate the coupling effect between the two charge sensors and their most directly coupled QDs in the QQD array as shown in Fig. \ref{figure:1}c and \ref{figure:1}d. In order to study the coupling effect related to LCS, a single QD (QD2) and LCS are tuned up as seen in the schematic sketch of Fig. \ref{figure:1}c. LCS and QD2 are represented by blue solid dots in the sketch and these two dots are directly coupled via the metal antenna wire CW1. Here, the rest four dots in the array are tuned off to ensure a conductive channel for QD2 and are displayed with dashed circles. The bottom panel of Fig. \ref{figure:1}c reveals the current through LCS $I_\mathrm{sdl}$ as a function of $V_\mathrm{gl2}$ and $V_\mathrm{G4}$. There is a high current line with a finite slope, which is a Coulomb peak of LCS when a level of LCS is aligned with the Fermi energy. The line has a finite slope due to a cross-talk effect between the two scanned gates. Notably, the high current line experiences sudden horizontal jumps at specific $V_\mathrm{G4}$ as marked by gray dotted lines. These jumps are due to the changes of the charge number (N) in QD2 with $V_\mathrm{G4}$ and the changes capacitively influence the electrochemical potential of LCS. In order to see the sensitivity of LCS on QD2, the gates are scanned along the blue dashed arrow (i.e., Path1) and corresponding current through LCS and QD2 are displayed in Fig. \ref{figure:1}e. The gates are scanned along Path1 to ensure a relatively high sensitivity of LCS in a large range of $V_\mathrm{G4}$ by compensating the cross-talk effect. In Fig. \ref{figure:1}e, we see that the current through LCS exhibits a sudden drop once a Coulomb peak appears in QD2 current. These correlated behaviors demonstrate that LCS can work as a charge sensor to probe the charge state of QD2. Here, the metal wire CW1 enhances the capacitive coupling between LCS and QD2, therefore ensuring a high charge sensitivity. Figure \ref{figure:1}d shows the coupling effect between QD5 and RCS. Similar as Fig. \ref{figure:1}c, only QD5 and RCS (blue solid dots) are tuned up and the other dots (dashed circles) in the array are tuned off. The bottom panel displays the capacitive coupling between QD5 and RCS, where the Coulomb peak of RCS has sudden shifts when the charge number (M) in QD5 varies. Then, gates are scanned along Path2 in the figure and corresponding measured results are shown in Fig. \ref{figure:1}f. Similarly, RCS is able to capture the charge variation in QD5 as a result of the enhanced coupling between the two dots via the metal wire CW2. As seen in Fig. \ref{figure:1}e and \ref{figure:1}f, when the dots in the array are tuned by gates, the gates to sensors should be adjusted accordingly to maintain an appropriate sensitivity for sensors. Such a linear combination of physical gates could generate so-called virtual gates, and the factors for the conversion is expressed with a matrix $M$, such that $ \mathrm{vG}_\mathrm{i}  = \sum_{j}^{} M_{ij}\cdot  \mathrm{G}_\mathrm{j} $, where i and j are the indices of the matrix $M$, G are physical gates and vG represent virtual gates. The strategy is employed to all of the rest experiments. In the following, multiple QDs in a linear array, up to a QQD, are defined and investigated.   

\begin{figure*}[!t] 
\centering
\includegraphics[width=1\linewidth]{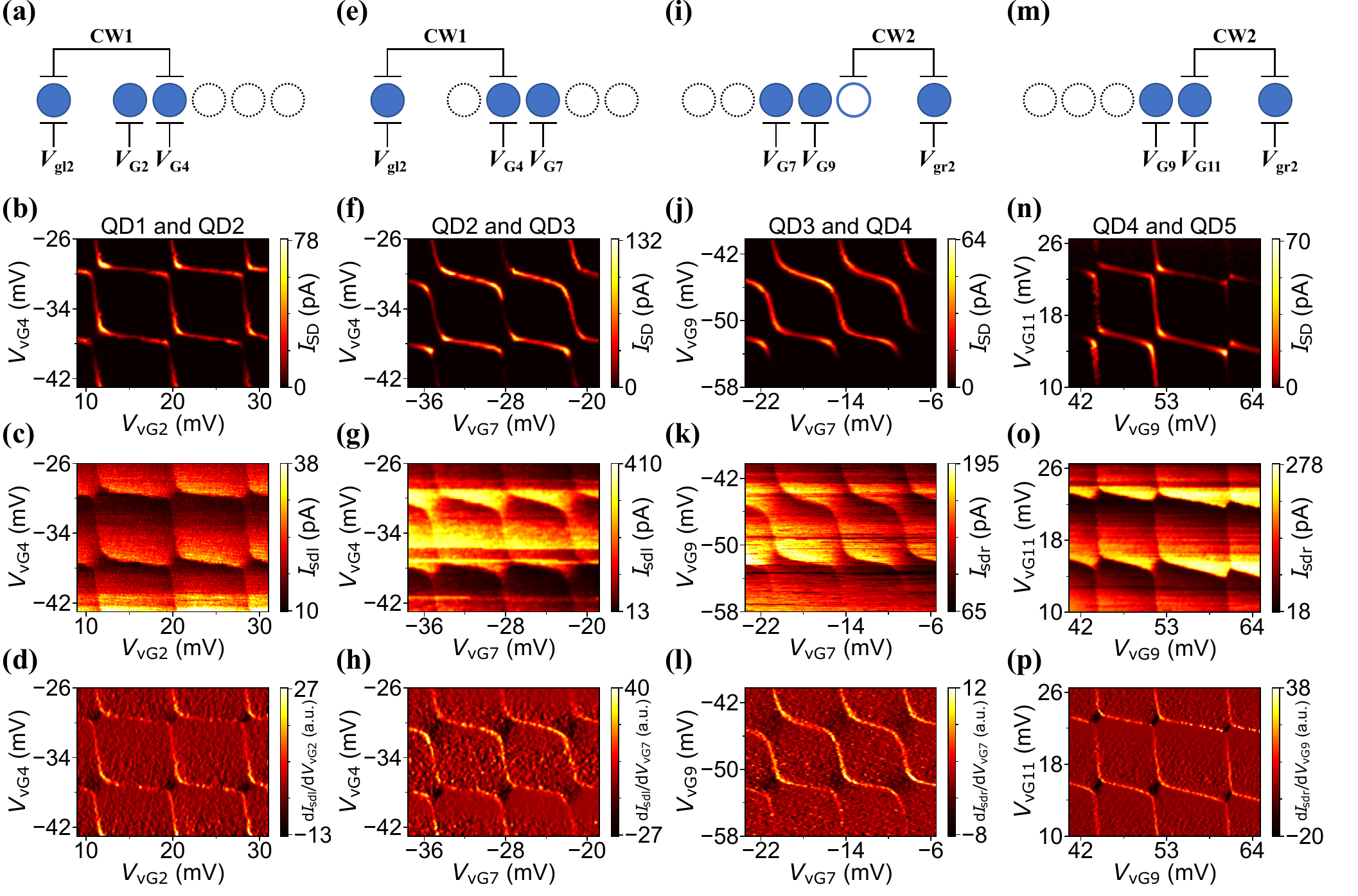}
\caption{\doublespacing \textbf{(a)}, \textbf{(e)}, \textbf{(i)} and \textbf{(m)} Schematic sketches of DQDs together with charge sensors in the case of QD1-QD2, QD2-QD3, QD3-QD4, and QD4-QD5, respectively. \textbf{(b)}, \textbf{(f)}, \textbf{(j)} and \textbf{(n)} Charge stability diagrams of the formed DQDs measured via direct current transport. \textbf{(c)}, \textbf{(g)}, \textbf{(k)} and \textbf{(o)} Charge stablity diagrams of the DQDs detected with charge sensors. \textbf{(d)}, \textbf{(h)}, \textbf{(l)} and \textbf{(p)} Transconductance of charge sensor signals.} \label{figure:2}
\end{figure*}

In order to validate the reproducibility and controllability of the device, the middle segment of the nanowire is configured into DQDs consecutively as shown in Fig. \ref{figure:2}. Figure \ref{figure:2}a presents a schematic sketch of a DQD (QD1 and QD2) together with LCS, and corresponding measurement results are displayed in Fig. \ref{figure:2}b-\ref{figure:2}d. During the measurements, virtual gate vG2 (vG4) is applied by considering the cross talk between physical gate G2 (G4) and the sensor gate gl2. Conversion matrices for the virtual gates are given in Section 2 of the Supporting Information. Figure \ref{figure:2}b shows the direct current transport through the DQD $I_\mathrm{SD}$ as a function of the virtual gate voltage $V_\mathrm{vG2}$ and $V_\mathrm{vG4}$, and the honeycomb patterns confirm the formation of a coupled DQD. The response of LCS, $I_\mathrm{sdl}$ versus $V_\mathrm{vG2}$ and $V_\mathrm{vG4}$, is presented in Fig. \ref{figure:2}c, and transconductance d$I_\mathrm{sdl}$/d$V_\mathrm{vG2}$ as a function of $V_\mathrm{vG2}$ and $V_\mathrm{vG4}$ is shown in Fig. \ref{figure:2}d. The charge sensor signals successfully manifest the boundaries between different charge states even in the case of inter-dot transition. In a similar way, coupled DQDs are constructed consecutively in other region of the middle segment in the nanowire. Figure \ref{figure:2}e-\ref{figure:2}h show the results of the DQD formed by QD2 and QD3, Fig. \ref{figure:2}i-\ref{figure:2}l present the results of the DQD formed by QD3 and QD4, and Fig. \ref{figure:2}m-\ref{figure:2}p are for the DQD made of QD4 and QD5. In the case of QD3 and QD4 (Fig. \ref{figure:2}i-\ref{figure:2}l), QD5 is also tuned up to keep a high sensitivity of RCS to QD3 and QD4 as the metal antenna wire CW2 is directly connected to QD5. In the above results, synchronized responses are seen between direct transport and charge sensor signals for all four DQDs. Particularly, charge sensor signal is clearly present even in the case that direct current transport is extremely weak. The effective charge sensing, especially in weak direct transport regimes, guarantees the investigation on the device in multiple QD configurations as follows. 

\begin{figure*}[!t] 
\centering
\includegraphics[width=1\linewidth]{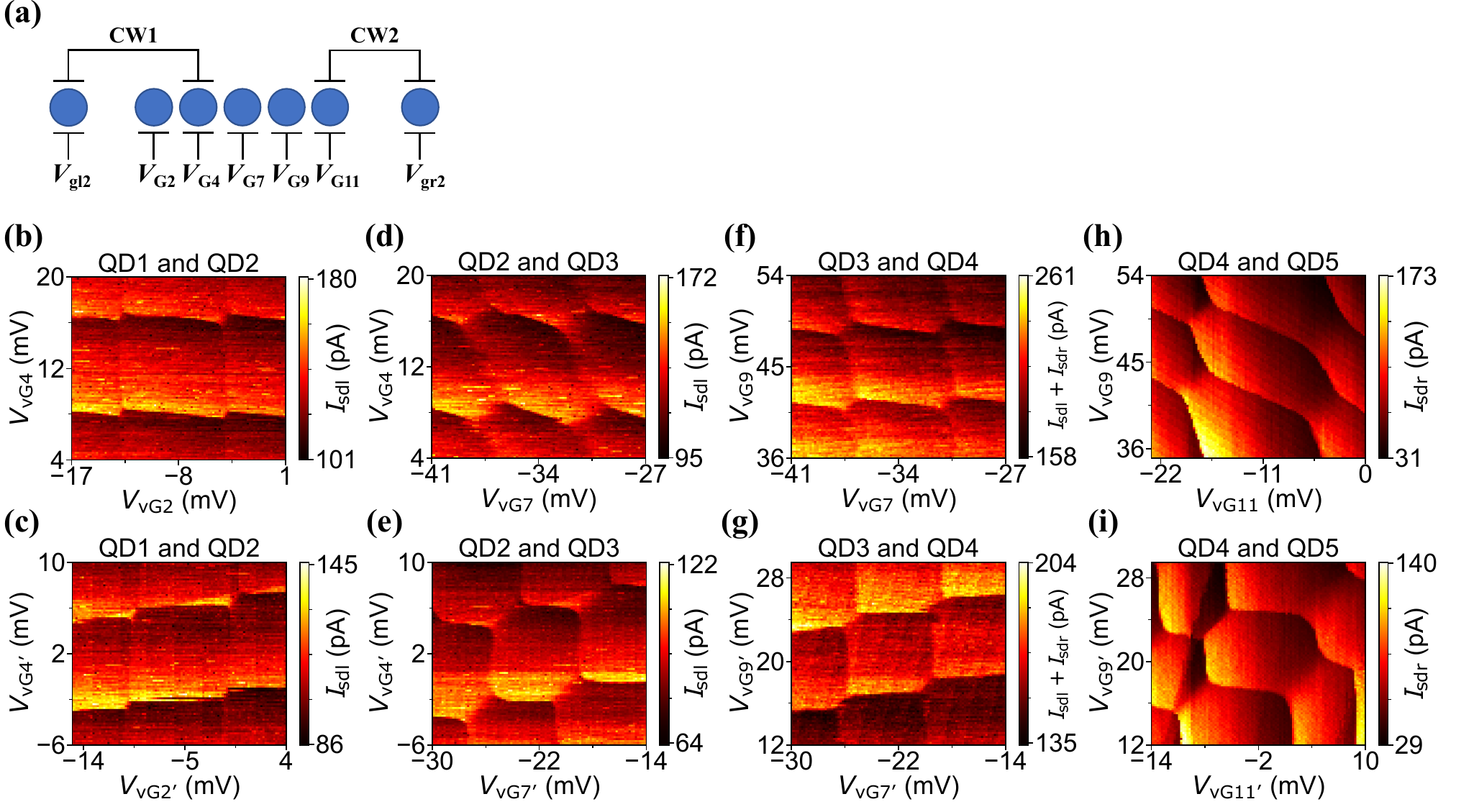}
\caption{\doublespacing \textbf{(a)} Schematic sketch of a QQD together with two charge sensors. \textbf{(b), (d), (f) and (h)} Charge stability diagrams by scanning virtual gates which
only compensate the cross talk between the QQD and charge sensors. \textbf{(c), (e), (g) and (i)} Charge stability diagrams by scanning virtual gates where the inter-dot capacitive couplings in the QQD are taken into account as well.}\label{figure:3}
\end{figure*}

Figure \ref{figure:3} shows the stability diagrams of a QQD detected solely by two charge sensor dots. As seen in the schematic sketch (Fig. \ref{figure:3}a), the QQD (QD1-QD5) is tuned up and two charge sensors are designated to detect the charge state properties of the QQD. The direct transport current through the QQD is hardly detectable and the charge characteristics of the QQD is only measurable with the two charge sensors. Figure \ref{figure:3}b, \ref{figure:3}d, \ref{figure:3}f and \ref{figure:3}h display the charge stability diagrams of the QQD, where each panel presents the charge states of two neighboring QDs. Here, virtual gates are converted from physical gates by considering the cross-talk effects between each plunger gate and charge sensor gates. The conversion matrix used in the measurements is provided in Section 2 in the Supporting Information. Note that when scanning the panel of QD3 and QD4 (Fig. \ref{figure:3}f) both charge sensors are energized to ensure a proper data visibility. In the figures, each two neighboring QDs exhibit a finite capacitive coupling as well as tunnel coupling. These inter-dot couplings in principle can be well tuned by barrier gates as demonstrated in our previous work\cite{WangXuming2021}. The plunger gate capacitance and inter-dot mutual capacitance are extracted from Fig. \ref{figure:3}b, \ref{figure:3}d, \ref{figure:3}f and \ref{figure:3}h and shown in Table S3 in the Supporting Information. Additionally, it is observed that the charge transition boundaries of each dot are not exactly horizontal or vertical as a result of finite cross talk between the plunger gates of each two neighboring QDs. In order to further prove the controllability on the device, a full compensation among the plunger gates for all QDs, including the QQD and charge sensors, is employed, and corresponding results are displayed in Fig. \ref{figure:3}c, \ref{figure:3}e, \ref{figure:3}g and \ref{figure:3}i. Here, the virtual gates are converted from seven physical plunger gates via a 7 by 7 matrix (see Section 2 in the Supporting Information). It is clear that the charge state boundaries of every QD in the QQD become horizontal or vertical, indicating that a nearly orthogonal space is formed within the virtual gates and the energy level of each dot can be tuned independently. Further measurements are done to validate the independence of the virtual gates in Section 4 in the Supporting Information. Notably, the data in Fig. \ref{figure:3}b-\ref{figure:3}i correspond to the same ranges of physical plunger gate voltages. The above results prove that the device is highly controllable and all QDs in the QQD array can be tuned simultaneously to a proper condition. Such a high-performance device is valuable for carrying out complex quantum tasks in nanowire systems.

\begin{figure*}[!t] 
\centering
\includegraphics[width=0.5\linewidth]{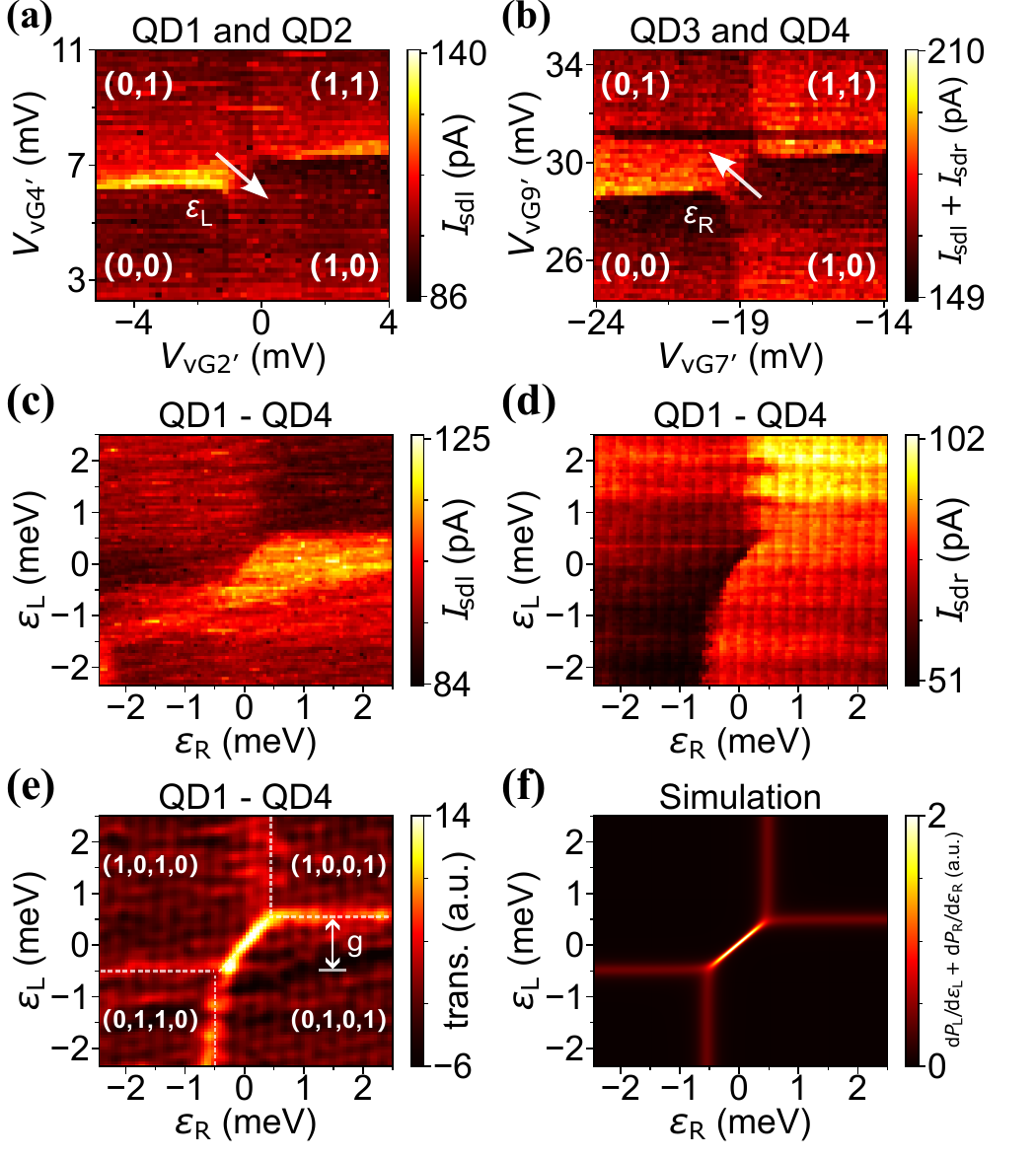}
\caption{\doublespacing \textbf{(a)} Stability diagram of the left DQD formed by QD1 and QD2. \textbf{(b)} Stability diagram of the right DQD composed of QD3 and QD4. In (a) and (b), the numbers in brackets represent the effective charge numbers in the dots, and the white arrows mark the detuning axis $\epsilon_\mathrm{L}$ and $\epsilon_\mathrm{R}$. \textbf{(c)} and \textbf{(d)} Stability diagram of the two coupled DQDs as a function of $\epsilon_\mathrm{L}$ and $\epsilon_\mathrm{R}$ detected by the left and right charge sensor, respectively. \textbf{(e)}  Stability diagram as a weighted sum of transconductance signals from the two charge sensors: $ a\cdot \mathrm{d}I_\mathrm{sdl}/\mathrm{d}\epsilon_\mathrm{L}+b\cdot \mathrm{d}I_\mathrm{sdr}/\mathrm{d}\epsilon_\mathrm{R}$, where $a = -139.2$ and $b = 116.1$. The arrow indicates the inter-DQD capacitive coupling $g\sim 242\,\mathrm{GHz}$. \textbf{(f)} Simulated charge stability diagram of two coupled DQDs based on a 4-dimensional Hamiltonian.}\label{figure:4}
\end{figure*}

It is known that two-qubit gates in spin and charge qubits strongly depend on the interactions between multiple dots, where exchange interaction and capacitive coupling are used respectively\cite{Nowack2011,Shulman2012,Li2015}. For instance, the speed of two-qubit gates in charge qubits is determined by the mutual capacitive interaction between the two qubits accommodated in two coupled DQDs. Here, we present an example of two strongly coupled DQDs within the QD array in Fig. \ref{figure:4}. Figure \ref{figure:4}a shows the charge stability diagram of the left DQD made of QD1 and QD2, and Fig. \ref{figure:4}b shows the result of the right DQD made of QD3 and QD4. The numbers in the brackets represent the effective charge numbers in the dots. The inter-dot tunnel coupling of the left (right) DQD is obtained to be $t_\mathrm{L}\,\sim 19.2\,\mathrm{GHz}$  ($t_\mathrm{R}\,\sim 19.7\,\mathrm{GHz}$) by fitting charge sensor response crossing the inter-dot transition (see details in Section 5 in the Supporting Information). In order to measure the capacitive coupling between the two DQDs, gates are scanned along the detuning axes, $\epsilon_\mathrm{L}$ and $\epsilon_\mathrm{R}$, as labelled in Fig. \ref{figure:4}a and \ref{figure:4}b. Corresponding charge stability diagrams as a function of $\epsilon_\mathrm{L}$ and $\epsilon_\mathrm{R}$ are shown in Fig. \ref{figure:4}c and \ref{figure:4}d, which are measured with LCS and RCS, respectively. Figure \ref{figure:4}e is obtained by summing the two differentiated sensor signals measured in Fig. \ref{figure:4}c and \ref{figure:4}d with a weight. In the figure, the numbers in the brackets indicate the effective charge numbers in the four dots. Horizontal and vertical bright lines represent a charge transfer within the left and right DQD, respectively. Interestingly, we see a substantial amount of capacitive interaction between the two DQDs. Corresponding phenomenon is that the charge state (0, 1, 1, 0) and (1, 0, 0, 1) are separated in the diagram due to Coulomb interaction. In a simple picture, the two effective charges are close in space for the state (0, 1, 1, 0) as compared to the state (1, 0, 0, 1), and therefore stronger Coulomb interaction exists for the former state. As a contrast, the states (1, 0, 1, 0) and (0, 1, 0, 1) have comparable Coulomb interaction and are hence adjacent to each other in the diagram. The capacitive interaction between the two DQDs can be directly extracted from the figure and the value is $g \sim 1\,\mathrm{meV}$ ($\sim242\,\mathrm{GHz}$). The coupling strength in our device is larger than those reported in GaAs- and Si-based QD arrays\cite{Li2015,Li2018,Samuel2019PRA}, implying the capability of accommodating ultra-fast two-qubit gate manipulations in the device. The coupling strength could also be calculated from a capacitive network of a QD array as developed in Ref \zcite{Samuel2019PRA}. By using the extracted capacitance values from experimental data (see Table S3 in the Supporting Information), the coupling strength of the two DQDs is calculated to be $1.021\,\mathrm{meV}$ (i.e., $247\,\mathrm{GHz}$), well consistent with the directly extracted value. Moreover, we take a 4-dimensional effective Hamiltonian model for a numerical simulation of two coupled DQDs (see details of the model in Section 6 of the Supporting Information). Figure \ref{figure:4}f displays the simulated stability diagram of two interacted DQDs. In the simulation, key parameters are adopted from the above experiments, where $t_\mathrm{L}\,\sim 19.2\,\mathrm{GHz}$, $t_\mathrm{R}\,\sim 19.7\,\mathrm{GHz}$ and $g \sim242\,\mathrm{GHz}$. It is clear that the simulated diagram agrees well the experimental data in Fig. \ref{figure:4}e, confirming the presence of strong inter-DQD interaction. Such a strong interaction supports fast multiple qubit state manipulations in nanowire-based QD devices.

In summary, a linear QD array, up to a QQD, integrated with two charge sensors is realized in an InAs nanowire via a fine finger-gate technique. In DQD configurations, the functionality of the charge sensors is validated by the synchronized signals from direct transport and charge sensors. Then, a QQD is tuned up with the assistance of the charge sensors and each dot in the array can be controlled independently by employing virtual gates. After that, four QDs in the array are selected to form two coupled DQDs, in which ultra-strong capacitive coupling between the DQDs is obtained. The coupling strength of the two DQDs is further confirmed with a numerical simulation based on a 4-dimensional Hamiltonian. We expect that the highly controllable one-dimensional QD array achieved in this work is valuable for constructing advanced quantum hardware in nanowire systems in the future.

\section{Author contributions}

H.Q.X conceived and supervised the project. D.P. and J.Z. grew the semiconductor InAs nanowire. Y.L. and S.Y. fabricated the device. J.-Y.W., W.L., Y.L., X.-F.L., H.G. and H.S. set up the measurement circuit in the dilution refrigerator. Y.L., X.-F.L. and J.-Y.W. performed the transport measurements. Z.-H.L. and Y.L. finished the simulation work. Y.L., X.-F.L., Z.-H.L., J.-Y.W. and H.Q.X. analyzed the measurement data. Y.L., J.-Y.W. and H.Q.X. wrote the manuscript with inputs from all the authors. 

\section{Supporting Information}
The Supporting Information is available free of charge.
The Supporting Information includes: details of material information,device fabrication,transport experiment setups and additional transport experiment data (PDF).

\begin{acknowledgement}
This work was supported by the National Natural Science Foundation of China (Grant Nos. 92165208, 12374480, 11874071, 92065106, 61974138, 12004039 and 12374459), the Ministry of Science and Technology of China through the National Key Research and Development Program of China (Grant Nos. 2017YFA0303304 and 2016YFA0300601). D.P. also acknowledges the supports from the Youth Innovation Promotion Association, Chinese Academy of Sciences (Nos. 2017156 and Y2021043). We would like to thank Po Zhang for technical support and Jiaan Qi for helpful discussions.
\end{acknowledgement}

\section{Data analysis and data availability} 

The raw data and the analysis files are available upon request. 
      
\section{Conflict of interests}
The authors declare no conflict of interests. 

\bibliography{references}

\newpage

\newpage

\newpage

\newpage


\end{CJK}
\end{document}


\begin{CJK}{UTF8}{gbsn} 

 \section{Section 1. Pinch-off voltages of the gates} 
 
 \par The transfer characteristics of individual gates are measured with $V_{bg}$ fixed at $2.5\,\mathrm{V}$ and the pinch-off voltages of all the top gates are in the range of $-1.5\,\mathrm{V}$ to $-0.2\,\mathrm{V}$ shown in Table S1, where G6 is not functional. In all measurements, G6 is floated. Based on the currents in the pinch-off state, all offsets from the circuit are also corrected.
\begin{table}[htbp]
    \centering
    \renewcommand{\thetable}{S1}
    \caption{The pinch-off voltages of all the top gates. (\textbf{$Unit: V$})}
    \begin{tabular}{ccccccccccc}
        \hline\hline\noalign{\smallskip}
        \textbf{G1} & \textbf{G2} & \textbf{G3} & \textbf{G4} & \textbf{G5} & \textbf{G6} & \textbf{G7} & \textbf{G8} & \textbf{G9} & \textbf{G10} & \textbf{G11}\\
        \noalign{\smallskip}\noalign{\smallskip}
        -1.410 & -1.230 & -1.180 & -1.300 & -1.120 & None & -1.120 & -1.123 & -1.200 & -1.000 & -1.490 \\
        \noalign{\smallskip}\hline\noalign{\smallskip}
        \textbf{G12} & \textbf{G13} & \textbf{PG1} & \textbf{gl1} & \textbf{gl2} & \textbf{gl3} & \textbf{PG2} & \textbf{gr1} & \textbf{gr2} & \textbf{gr3} \\
        \noalign{\smallskip}\noalign{\smallskip}
        -1.270 & -1.370 & -0.550 & -0.680 & -0.700 & -0.600 & -0.540 & -0.340 & -0.280 & -0.280 \\ 
        
        \noalign{\smallskip}\hline\hline
    \end{tabular}
\end{table}

\section{Section 2. Conversion matrices for virtual gates} 

In all virtual gates, the voltage values of barrier gates are always fixed within the target working areas, so the barrier gate is not written into the virtual gate matrices (VGMs). When measuring QDi-QD(i+1) (i=1, 2) and LCS (Fig. 2b-2d and Fig. 2f-2h), the VGMs in the basis of two plunger gates of double dot and gl2, i.e. $\Big\{G2,G4,gl2\Big\}$ and $\Big\{G4,G7,gl2\Big\}$, are:
 \begin{align}
M^{}_{QD1-2\,with\,LCS} =     \left(\begin{array}{ccc}
 1 & 0 & 0\\ 0 & 1 & 0.157\\ 0.007 & 0.051 & 1\\
\end{array}\right),\,
M^{}_{QD2-3\,with\,LCS} =     \left(\begin{array}{ccc}
 1 & 0 & 0.102\\ 0 & 1 & 0\\ 0.052 & 0.003 & 1\\
\end{array}\right)\label{Cd-l}  
 \end{align}
When measuring QDi-QD(i+1) (i=3, 4) and RCS (Fig. 2j-2l and Fig. 2n-2p), the VGMs in the basis of two plunger gates of double dot and gr2, i.e. $\Big\{G7,G9,gr2\Big\}$ and $\Big\{G9,G11,gr2\Big\}$, are:
 \begin{align}
M^{}_{QD3-4\,with\,RCS} =     \left(\begin{array}{ccc}
 1 & 0 & 0\\ 0 & 1 & 0.077\\ 0.004 & 0.080 & 1\\
\end{array}\right),\,
M^{}_{QD4-5\,with\,RCS} =     \left(\begin{array}{ccc}
 1 & 0 & 0\\ 0 & 1 & 0.131\\ 0.079 & 0.065 & 1\\
\end{array}\right)\label{Cd-l}  
 \end{align}
When measuring the QQD with two charge sensors (Fig. 3b, 3d, 3f and 3h), the VGM in the basis of $\Big\{G2,G4,G7,G9,G11,gl2,gr2\Big\}$ for the virtual gates which only compensates for the shift of charge sensors is:
 \begin{align}
M^{}_{QQD} =     \left(\begin{array}{ccccccc}
 1 & 0 & 0 & 0 & 0 & 0.020 & 0\\ 0 & 1 & 0 & 0 & 0 & 0.143 & 0\\ 0 & 0 & 1 & 0 & 0 & 0.020 & 0\\ 0 & 0 & 0 & 1 & 0 & 0 & 0.050\\ 0 & 0 & 0 & 0 & 1 & 0 & 0.125\\ 0.005 & 0.052 & 0.005 & 0 & 0 & 1 & 0\\ 0 & 0 & 0.004 & 0.070 & 0.070 & 0 & 1\\
\end{array}\right),\label{Cd-l}  
 \end{align}
and that for the virtual gates which extra consider the inter-dot coupling in QQD (Fig. 3c, 3e, 3g and 3i) is:
 \begin{align}
 M^{}_{QQD^{’}} =     \left(\begin{array}{ccccccc}
 1 & 0.091 & 0 & 0 & 0 & 0.020 & 0\\ 0.160 & 1 & 0.290 & 0 & 0 & 0.143 & 0\\ 0 & 0.280 & 1 & 0.111 & 0 & 0.020 & 0\\ 0 & 0 & 0.210 & 1 & 0.350 & 0 & 0.050\\ 0 & 0 & 0 & 0.270 & 1 & 0 & 0.125\\ 0.005 & 0.052 & 0.005 & 0 & 0 & 1 & 0\\ 0 & 0 & 0.004 & 0.070 & 0.070 & 0 & 1\\
\end{array}\right).\label{Cd-l}  
\end{align}

\clearpage

\section{Section 3. Capacitance extraction of the device} 
\par Figure S1 shows the Coulomb diamonds of the single quantum dot QD1 and QD5. From the figure, we can extract capacitance between quantum dot and source/drain electrode $C_{S}$/$C_{D}$.\cite{Hanson2007} Extracted values are listed in Table S3.

\begin{figure*}[!h]
\centering
\includegraphics[width=1\linewidth]{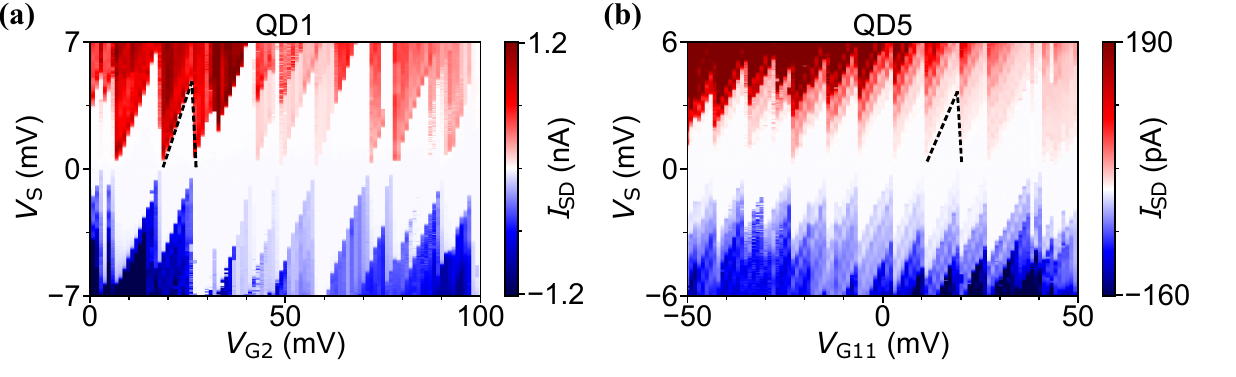} 
\caption{\doublespacing \textbf{Coulomb diamonds of single quantum dots.} \textbf {(a)} Coulomb diamond of QD1. \textbf {(b)} Coulomb diamond of QD5. The black dashed lines indicate the slope used to extract the capacitance between quantum dot and source/drain electrode.
}\label{figure:S1}
\end{figure*}

\par To make the formula easier to be read, we use Pi (i=1,2,3,4,5) to represent the plunger gate of each quantum dot, where $P1=G2, P2=G4, P3 =G7, P4=G9$ and $P5=G11$.
The capacitance $C_{Pi}$ between QDi and their own plunger gate Pi (i=1,2,3,4,5)
could be calculated by
\begin{align}
C_{Pi}=\frac{e}{\Delta V_{Pi}}, \,i=1,2,3,4,5, 
\label{CPi}
\end{align}
where $e$ is the elementary charge, $\Delta V_{Pi}$ is the current peak spacings of QDi (i=1,2,3,4,5). The inter-dot capacitance ($C_{Mij}$) between QDi and QDj ($i,j=1,2,3,4,5$ and $i\neq j$) is expressed as\cite{Hanson2007}
\begin{align}\begin{array}{c}
C_{M12}=\frac{\Delta V^{M12}_{P2}}{\Delta V_{P2}}(C_{S}+C_{P1}+C_{M12}),\,\,\,\,\,\, \\
C_{M12}=\frac{\Delta V^{M12}_{P1}}{\Delta V_{P1}}(C_{M12}+C_{P2}+C_{M23}), \\
C_{M23}=\frac{\Delta V^{M23}_{P3}}{\Delta V_{P3}}(C_{M12}+C_{P2}+C_{M23}), \\
C_{M23}=\frac{\Delta V^{M23}_{P2}}{\Delta V_{P2}}(C_{M23}+C_{P3}+C_{M34}), \\
C_{M34}=\frac{\Delta V^{M34}_{P4}}{\Delta V_{P4}}(C_{M23}+C_{P3}+C_{M34}), \\
C_{M34}=\frac{\Delta V^{M34}_{P3}}{\Delta V_{P3}}(C_{M34}+C_{P4}+C_{M45}), \\
C_{M45}=\frac{\Delta V^{M45}_{P5}}{\Delta V_{P5}}(C_{M34}+C_{P4}+C_{M45}), \\
C_{M45}=\frac{\Delta V^{M45}_{P4}}{\Delta V_{P4}}(C_{M45}+C_{P5}+C_{D}),\,\,\,\,\,\, \\
\end{array}
\label{CM}
\end{align}
where the current line shift $\Delta V^{Mij}_{Pi/j}$ ($i,j=1,2,3,4,5$ and $i\neq j$) and the current peak spacings $\Delta V_{Pi}$ are got in Fig. 3(b-e). Taking Fig. 3b as an example, we label $\Delta V^{M12}_{P1/2}$ and $\Delta V_{P1/2}$, as shown in Fig. S2. Other data extraction is similar to this. Extracted capacitance values are shown in Table S3.

The energy level arm factors of QD1/QD5 could be obtained from the Coulomb diamonds in Fig. S1. The energy level arm factors of the other quantum dots can be derived sequentially from the slopes of the inter-dot transition lines in the stability diagrams  shown in Fig. 3b, 3d, 3f and 3h.\cite{Hensgens2017,Unseld2023} In our device, the arm factors of the plunger gates for the five quantum dots in the QQD are considered approximately equal, which are taken as about 0.52 meV/mV.

\clearpage

\begin{figure*}[!h]
\centering
\includegraphics[width=0.5\linewidth]{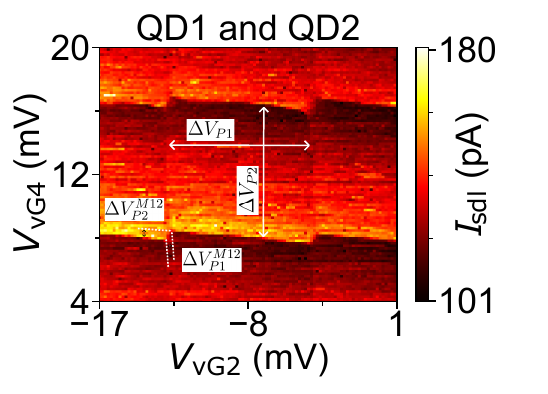} 
\caption{\doublespacing \textbf{Parameter extraction of capacitance model.} The current line shift $\Delta V^{M12}_{P1/2}$ and the current peak spacings $\Delta V_{P1/2}$ are labeled for Fig. 3b as an example.
}\label{figure:S2}
\end{figure*}

\begin{table}[htbp]
    \centering
    \renewcommand{\thetable}{S3}
    \caption{Extracted capacitances of the device. (\textbf{$Unit: aF$})}
    \begin{tabular}{ccccccccccc}
        \hline\hline\noalign{\smallskip}
        \textbf{$C_{S}$} & \textbf{$C_{D}$} & \textbf{$C_{G2}$} & \textbf{$C_{G4}$} & \textbf{$C_{G7}$} & \textbf{$C_{G9}$} & \textbf{$C_{G11}$} & \textbf{$C_{M12}$} & \textbf{$C_{M23}$} & \textbf{$C_{M34}$} & \textbf{$C_{M45}$}\\
        \noalign{\smallskip}\noalign{\smallskip}
        5.080 & 17.333 & 18.263 & 19.753 & 25.000 & 22.535 & 13.008 & 1.134 & 6.624 & 2.890 & 8.202 \\
        
        \noalign{\smallskip}\hline\hline
    \end{tabular}
\end{table}

\section{Section 4. Verification on the independence of the virtual gates in QQD configuration} 
\par In this section, we experimentally verify the independence of the virtual gates in QQD configuration. Figure S3 shows corresponding experimental results. Here, we scan charge stability diagrams with two virtual plunger gates while varying the voltage of another virtual gate. For instance, Fig. S3a-S3c display charge stability diagrams by scanning virtual gate vG2$'$ and vG4$'$ at three different voltages of vG7$'$. The rest panels in Fig. S3 testify the independence of the other virtual gates. We see that the cross-talk effect is much alleviated by using virtual gates despite that the effect is not perfectly eliminated. 

\begin{figure*}[!h]
\centering
\includegraphics[width=1\linewidth]{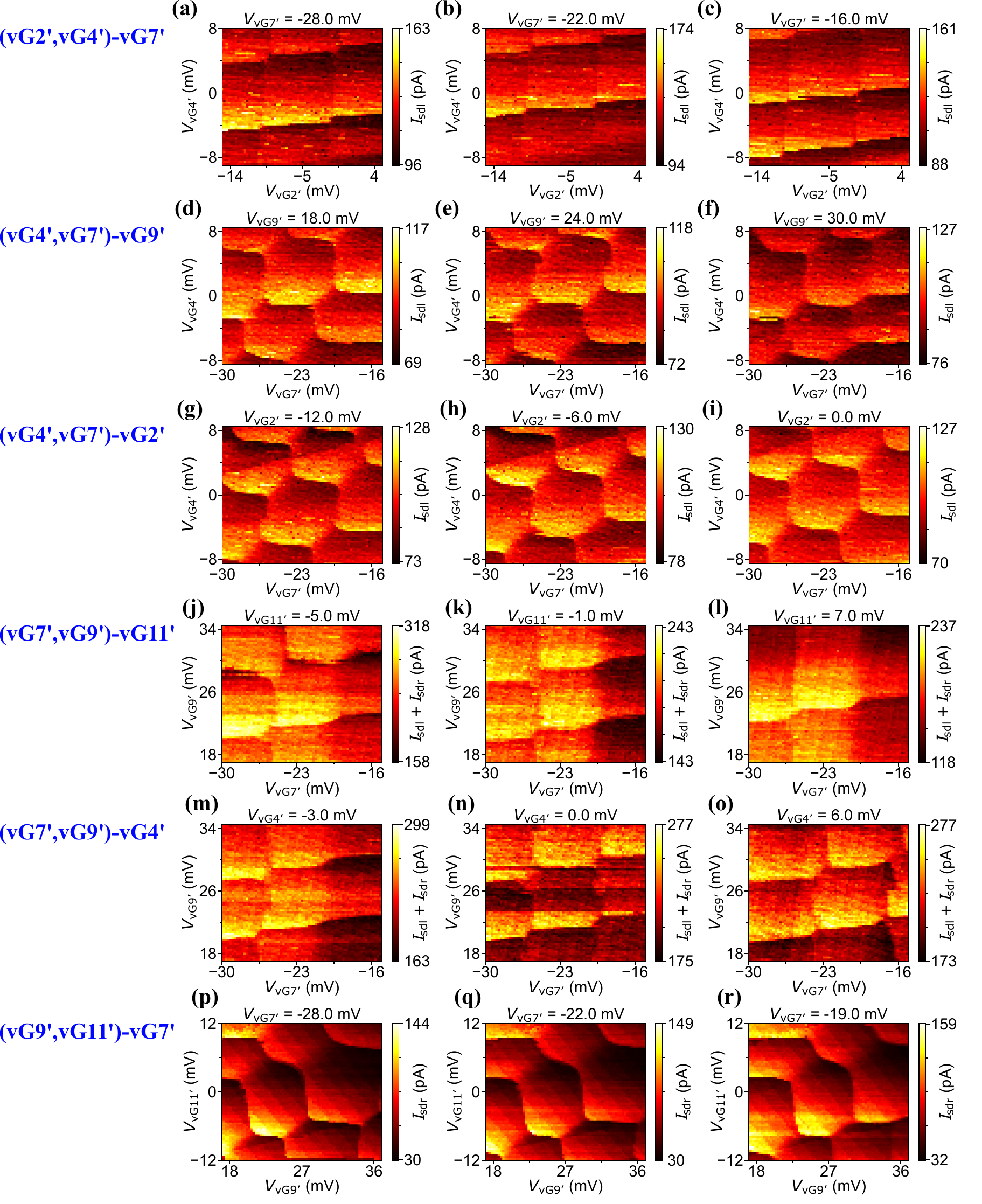} 
\caption{\doublespacing \textbf{Independence test on virtual gates in QQD.} Two-dimensional charge stability diagrams by scanning the virtual gates for QD1-QD2 at different voltages of vG7$'$ \textbf{(a-c)}, for QD2-QD3 at different voltages of vG9$'$ \textbf{(d-f)} and vG2$'$ \textbf{(g-i)}, for QD3-QD4 at different voltages of vG11$'$ \textbf{(j-l)} and vG4$'$ \textbf{(m-o)}, and for QD4-QD5 at different voltages of vG7$'$ \textbf{(p-r)}.
}\label{figure:S3}
\end{figure*}

\clearpage
\section{Section 5. Fitting inter-dot tunnel couplings in DQDs} 
\par Figure S4a and S4b show the response of the charge sensor signals while scanning gate voltages along detuning axes for the left DQD (QD1-QD2) and the right DQD (QD3-QD4), respectively. Here, the experimental data (red dots) are obtained from the line-cut across the polarization line of inter-dot charge transport near the gate voltage regions shown in Fig. 4a and Fig. 4b by using the "qtplot" toolkit (\href{https://github.com/Rubenknex/qtplot}{https://github.com/Rubenknex/qtplot}).

Inter-dot tunnel coupling can be extracted from the formula $I^{}_{sig}=\frac{\epsilon^{}_{}}{\sqrt[]{\epsilon^{2}_{}+4t^{2}_{}}}tanh
\frac{\sqrt[]{\epsilon^{2}_{}+4t^{2}_{}}}{2k^{}_{B}T^{}_{e}}$\cite{DiCarlo2004,Unseld2023}, where $\epsilon$ is detuning energy, $t$ is tunnel coupling, $T^{}_{e}$ is electron temperature and $k^{}_{B}$ is Boltzmann constant. Here, $T^{}_{e}$ is set to be $60\,\mathrm{mK}$. The blue traces in Fig. S4 are fitting curves to the data by using the above formula. The inter-dot tunnel coupling of the left DQD is extracted to be $t^{}_{L}=19.2\,\mathrm{GHz}$ and that of the right DQD is $t^{}_{R}=19.7\,\mathrm{GHz}$.

\begin{figure*}[!h]
\centering
\includegraphics[width=1\linewidth]{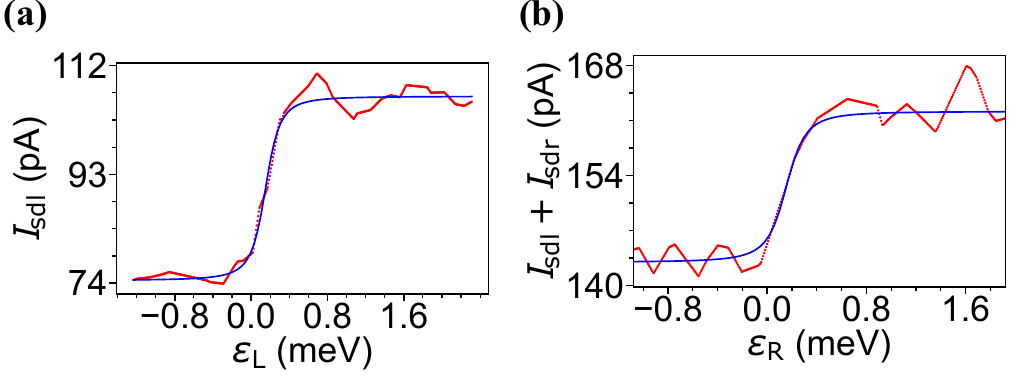} 
\caption{\doublespacing Extracting inter-dot tunnel couplings for the left DQD \textbf {(a)}, and the right DQD \textbf{(b)}. Red dots are experimental data and blue traces are fitting curves. $t_{L}$ is $\sim19.2\,\mathrm{GHz}$ and $t_{R}$ is $\sim19.7\,\mathrm{GHz}$. 
}\label{figure:S4}
\end{figure*}

\section{Section 6. Simulation } 
\par 

In order to capture the stability diagram of the two coupled DQDs (Fig.4(e)), we employ a 4-dimensional effective Hamiltonian
 \begin{align}
 H=\frac{\epsilon^{}_{L}}{2} \sigma^{L}_{z}\otimes I^{R}_{} +t^{}_{L}\sigma^{L}_{x}\otimes I^{R}_{} +\frac{\epsilon^{}_{R}}{2}I^{L}_{}\otimes \sigma^{R}_{z}+t^{}_{R} I^{L}_{}\otimes \sigma^{R}_{x}+\frac{g}{4} ( I^{L}_{}-\sigma^{L}_{z}) \otimes (I^{R}_{}-\sigma^{R}_{z})
 \end{align}
where $\epsilon^{}_{L/R}$ and $t^{}_{L/R}$ represent the detuning and the inter-dot tunneling amplitudes of the left and right DQD, $I^{L/R}_{}$ and $\sigma^{L/R}_{x, z}$ are Pauli matrices defined in the 2-dimensional charge space of the two DQDs, and $g$ specifies the capacitive coupling between them. By introducing the orthonormal basis of $\{|0^{}_{12},0^{}_{34}\rangle ,  |0^{}_{12},1^{}_{34}\rangle,  |1^{}_{12},0^{}_{34}\rangle, |1^{}_{12},1^{}_{34}\rangle \}$, in which  $|n^{}_{12}, m^{}_{34}\rangle= |n^{}_{12}\rangle \otimes|m^{}_{34}\rangle $  with $n,m=0,1$, $\sigma^{L/R}_{z}|0^{}_{12/34 }\rangle= |0^{}_{12/34}\rangle $ and $\sigma^{L/R}_{z}|1^{}_{12/34 }\rangle=-|1^{}_{12/34 }\rangle$, the Hamiltonian can be expanded as
\begin{align}
 H^{}_{\rm  }=\left[\begin{array}{cccc}
 \frac{\epsilon^{}_{L}}{2}+\frac{\epsilon^{}_{R}}{2} & t^{}_{R} &   t^{}_{L}& 0\\
 t^{}_{R}&  \frac{\epsilon^{}_{L}}{2}-\frac{\epsilon^{}_{R}}{2}& 0 &t^{}_{L}\\
 t^{}_{L}&0&  -\frac{\epsilon^{}_{L}}{2}+\frac{\epsilon^{}_{R}}{2}& t^{}_{R}\\
 0 &t^{}_{L}& t^{}_{R} & - \frac{\epsilon^{}_{L}}{2}-\frac{\epsilon^{}_{R}}{2}+g
 \end{array}\right]
 \end{align}
Then, the energy eigenvalues and eigenstates of the two coupled DQDs can be obtained through direct diagonalization of the matrix Hamiltonian. Specifically, let $|\Psi^{}_{j=1-4}\rangle$ denote the eigenstates which can be expressed by
  \begin{align}
  |\Psi^{}_{j}\rangle = \nu^{}_{j,1}|0^{}_{12}, 0^{}_{34}\rangle + \nu^{}_{j,2}|0^{}_{12}, 1^{}_{34}\rangle + \nu^{}_{j,3}|1^{}_{12}, 0^{}_{34}\rangle + \nu^{}_{j,4}|1^{}_{12}, 1^{}_{34}\rangle\ ,
  \label{PSW}
  \end{align}
 with $\nu^{}_{j,1-4}$ representing the combination coefficients.
Based on this, the charge polarization of the left/right DQD can be derived as\cite{DiCarlo2004,Samuel2019PRA}
  \begin{align}
  P^{}_{L /R }  =\frac{1}{ Z} \sum^{4}_{j=1} \langle \Psi^{}_{j}|\sigma^{L/R }_{z} |\Psi^{}_{j}\rangle  e^{-\frac{E^{}_{j}}{k^{}_{\rm B} T^{}_{e}}}_{}
  \end{align}
  where $Z= \sum^{}_{j} e^{-E^{}_{j}/(k^{}_{\rm B} T^{}_{e}) }_{}$ is the partition function, $E^{}_{j}$ is the energy eigenvalues, $k^{}_{\rm B}$ is the Boltzmann constant and $T^{}_{e}$ is the electron temperature which is set to be $60\,\mathrm{mK}$. By using the explicit form of the energy states in Eq.~(\ref{PSW}), the polarizations can be further expanded as
  \begin{align}
  P^{}_{L}=& \frac{1}{Z} \sum^{4}_{j=1}\left(|\nu^{}_{j,1}|^{2}_{}+|\nu^{}_{j,2}|^{2}_{}-|\nu^{}_{j,3}|^{2}_{}-|\nu^{}_{j,4}|^{2}_{} \right)e^{-\frac{E^{}_{j}}{k^{}_{\rm B} T^{}_{e}}}_{}\nonumber\\
  P^{}_{R}=&\frac{1}{Z} \sum^{4}_{j=1}\left(|\nu^{}_{j,1}|^{2}_{}-|\nu^{}_{j,2}|^{2}_{}+|\nu^{}_{j,3}|^{2}_{}-|\nu^{}_{j,4}|^{2}_{} \right)e^{-\frac{E^{}_{j}}{k^{}_{\rm B} T^{}_{e}}}_{}\ .
 \end{align}
 
Indeed, the detected stability diagram is correlated to the derivative of the two polarizations, i.e., $f=\frac{d P^{}_{L} }{d  \epsilon^{}_{L}}+\frac{d P^{}_{R} }{d  \epsilon^{}_{R}}$, because the transfer of an electron between the QDs can be reflected by the emergence of a peak in $f$. To be more specific, Fig. 4f shows the numerical distribution of the derivative factor $f$ in $\epsilon^{}_{L}-\epsilon^{}_{R}$ plane by adopting parameters extracted from experiments, $t^{}_{L}=19.2\,\mathrm{GHz}$, $t^{}_{R}=19.7\,\mathrm{GHz}$ and $g^{}_{}=242\,\mathrm{GHz}$. The simulated diagram (Fig. 4f) is well consistent with the experimental results shown in Fig. 4e.

\clearpage
\bibliography{references}

\end{CJK}